\documentclass[11pt, a4paper]{article}
\usepackage[utf8]{inputenc}
\usepackage[T1]{fontenc}
\usepackage{graphicx}  
\usepackage{amsmath, amssymb} 
\usepackage{geometry} 
\usepackage{hyperref} 
\usepackage{lineno}   
\usepackage{setspace} 
\usepackage{subcaption}
\usepackage{authblk}
\usepackage{threeparttable}
\usepackage{booktabs}
\usepackage{multirow}
\hypersetup{hidelinks}

\newcommand{\E}{\mathbb E}

\newcommand{\dd}{\,\mathrm d}
\newcommand{\kBT}{k_{\mathrm B}T}

\newcommand{\M}{\mathsf M}

\title{Unsupervised Thermodynamics of Molecular Diffusion Models: Action-Operator Semantics and Auditable Free-Energy Readout}
\author{
    Wenjie Xi$^{1}$\\
    \normalsize{$^{1}$Department of Physics and HK Institute of Quantum Science \& Technology, The University of Hong Kong, Pokfulam Road, Hong Kong, China}\\
    \normalsize{E-mail: wjxi@connect.hku.hk}
}

\date{\today}

\begin{document}

\maketitle

\begin{abstract}

Diffusion models are increasingly utilized for modeling molecular structures and conformational ensembles, yet the thermodynamic meaning of their learned representations 
and scores remains elusive. 
To resolve this ambiguity, we introduce a mathematically 
consistent action-operator framework natively compatible with diffusion models. 
By defining a fixed molecular environment as a base action $S_0(x)$ and an alchemical perturbation as an operator $O(x)$, standard diffusion noising induces effective noised actions and operators whose gradients and alchemical derivatives are directly represented 
by the model's learned fields. 
This rigorous self-consistency enables a ``noisy operator 
bridge'' capable of reading out free-energy differences ($\Delta F$) from endpoint ensembles and per-frame evaluations. 
In controlled experiments on alanine dipeptide systems, we show that incorporating physical inductive biases enables partial recovery of the base action and perturbation operator. 
When applied to a challenging C6-H to C6-F ligand-pocket nonbonded perturbation (185L/IND) with negligible phase-space overlap, our supervised bridge estimates the alchemical $\Delta F$ within approximately $1\ k_\mathrm{B}T$ of a stable 19-state MBAR reference. 
Finally, we demonstrate that endpoint coordinates and binary labels alone are sufficient to partially recover the operator shape and a centered free-energy scale 
without any force or action supervision. 
This work provides a rigorous path toward transforming generative molecular diffusion models from black-box coordinate samplers into auditable thermodynamic estimators.

\end{abstract}

\section{Introduction}

Diffusion models have emerged as a powerful paradigm for representation and simulation of molecular structures and conformational ensembles~\cite{sohldickstein2015,ho2020,song2021,arts2023}.
While these models are typically evaluated on the physical plausibility of their generated coordinates, such spatial metrics leave their underlying thermodynamic ensembles underdetermined.
Though recent studies have begun to investigate how diffusion scores correspond to physical quantities~\cite{plainer2025consistent,mate2024,sarma2025extract,xie2026enhanced,xi2026auto}, establishing a mathematically consistent framework that connects local scores directly to auditable thermodynamic ensembles remains highly desirable.

To bridge this gap, we propose an action-operator interpretation for molecular diffusion models that is mathematically and structurally consistent with standard diffusion frameworks. 
The core physical intuition is that a declared molecular environment defines a reduced base action $S_0(x)$, while a chemical alteration—such as a ligand condition or an alchemical mutation—defines a perturbation operator $O(x)$ that shifts the system's action to $S_\kappa(x) = S_0(x) + \kappa O(x)$ along an auxiliary path coordinate $\kappa$. 
The normalized forward diffusion kernel naturally maps these clean actions to noise-dependent effective actions, such that the model's learned score fields correspond exactly to the gradients of these noised actions when equipped with appropriate inductive biases. 
Simultaneously, the alchemical derivatives of these effective actions yield the conditional expectations of the operators.

Under this semantics, molecular configurations generated by the model during generalization are rendered thermodynamically auditable, enabling us to assess whether the underlying $S_0$ and $O$ representations are physically sound. 
Crucially, a key thermodynamic consequence of this formulation is that normalized noising preserves the partition function ($Z_{\kappa,\sigma} = Z_\kappa$).
This invariant allows us to construct a ``noisy operator bridge'' to read out absolute free-energy differences ($\Delta F$) through a path integral of the effective operator in noise space.

In this work, we carry out four controlled experiments to evaluate this action-operator semantics sequentially, establishing a clear path from representation learning to thermodynamic diagnostics:
First, we investigate whether diffusion-style neural networks can accurately capture and represent the base-action $S_0$ of a fixed molecular ensemble (alanine dipeptide), and demonstrate how this representation scales with model capacity and data size.
Second, we fix $S_0$ and evaluate whether the condition-dependent perturbation operator $O$ can be recovered from biased ensembles using cross-entropy classification and clean score matching.
While complete recovery remains challenging, the recovery accuracy scales positively with larger dataset sizes and model
capacities.
Third, we apply the fully integrated noisy operator bridge to a challenging $\mathrm{C6}\text{-}\mathrm{H}$ to $\mathrm{C6}\text{-}\mathrm{F}$ ligand-pocket nonbonded perturbation in 185L/IND characterized by negligible endpoint phase-space overlap.
Given accurate evaluations of the base action and operator, we show that the bridge estimates the alchemical free-energy difference as $-24.74 \pm 0.10~k_{\mathrm{B}}T$, which is within $1.08 \pm 0.10~k_{\mathrm{B}}T$ of a stable 19-state MBAR reference ($-25.82~k_{\mathrm{B}}T$).
Finally, in an endpoint-only gauge audit (Experiment 4A), we demonstrate that combining a frozen base representation with endpoint thermodynamic consistency can identify the 
nonconstant operator shape and a gauged physical free-energy scale purely from coordinate distributions and binary labels, completely bypassing physical action or force supervision.

This approach bypasses the fundamental statistical bottleneck of traditional free-energy perturbation (FEP) and thermodynamic integration (TI) methods, which typically require simulating a large number of intermediate alchemical states to overcome poor phase-space overlap~\cite{zwanzig1954,kirkwood1935,bennett1976,shirts2008,meng1996,jarzynski1997,crooks1999}.
While advanced transport-based estimators, such as LBAR, Boltzmann Generators, Neural TI, and flow-matching methods, improve endpoint overlap by learning invertible molecular mappings or interpolating intermediate densities~\cite{jarzynski2002,wirnsberger2020,yoo2023,noe2019,ding2021,mate2023,mate2024,mate2025,freeflow2025,feat2025}, our noisy bridge represents a conceptually distinct paradigm. 
It does not compute change-of-variables Jacobians or generate nonequilibrium work trajectories. 
We make the methodological boundaries explicit: while the supervised bridge utilizes endpoint ensembles combined with declared per-frame reduced-action evaluations, the 
endpoint-only audit demonstrates that coordinate distributions alone can identify a relative free-energy gauge without online energy supervision. 
The key contribution of our Action-Operator semantics is demonstrating how a generative diffusion architecture can be systematically white-boxed to deliver auditable thermodynamic readouts.

\section{Fixed Environment as a Base Action \texorpdfstring{\(S_0\)}{S0}}

\subsection{Physical Picture}

At a fixed temperature and molecular environment, an equilibrium molecular ensemble in coordinate space $x$ residing on a Riemannian manifold $\mathcal{M}$ is characterized by the probability measure:
\begin{equation}
    p_0(x)=Z_0^{-1}e^{-S_0(x)}.
\end{equation}
For an all-atom classical force field with potential energy \(U_0(x)\), this action corresponds to the dimensionless, reduced potential \(S_0(x)=\beta U_0(x)\). 
A diffusion model trained on a fixed environment does not merely fit an untyped distribution of coordinates, but rather learns noised information about \(S_0\) through a normalized noising kernel \(K_\sigma(z\mid x)\). 
This process induces a noised density \(p_{0,\sigma}(z) = \int K_\sigma(z\mid x)p_0(x)\dd x\) and a corresponding effective noised action \(S_{0,\sigma}(z)=-\log p_{0,\sigma}(z)+\mathrm{const}\). 
For constrained or mass-metric coordinates, this effective action defines a natural score vector field:
\begin{equation}
v_{0,\sigma}(z) = \M^{-1}\nabla_z\log p_{0,\sigma}(z) = -\M^{-1}\nabla_zS_{0,\sigma}(z),
\end{equation}
where \(\M\) is the symmetric, positive-definite mass matrix (or the Riemannian metric tensor of the constraint manifold), \(\nabla_z\) is the Euclidean gradient, and the natural score \(v_{0,\sigma}(z)\) is transformed from the Euclidean score \(\nabla_z\log p_{0,\sigma}(z)\) via the metric inverse \(\M^{-1}\). This formulation establishes the noised action \(S_{0,\sigma}\) as a native, integrable diffusion quantity.

\subsection{Model architecture}

For semantic readout, we enforce score integrability by parameterizing a componentized scalar-action network \(A_\theta(z,\sigma)\). 
This network receives molecular features (atom identities, distances, local internal coordinates, and noise embeddings) and is assembled from topology-conditioned component heads—specifically bond, angle, torsion, 1--4 exception, and direct nonbonded terms in our ALA2 implementation.
Rather than employing an independent vector head, the score is 
obtained via automatic differentiation of this assembled action:
\begin{equation}
\label{eq:score_autodiff}
v_\theta(z,\sigma) = -\M^{-1}\nabla_z A_\theta(z,\sigma),
\end{equation}
followed by tangent projection when molecular constraints or mass-metric coordinates are used. 
This architecture ensures that \(A_\theta(z,\sigma)\) directly represents the effective noised action up to a \(z\)-independent additive constant, establishing the necessary mathematical gauge for downstream auditing.

For clean samples \(x\sim p_0\), we define noised positions on the constraint manifold via tangent noise and retraction:
\begin{equation}
\label{eq:retraction_noise}
z=\mathrm{Retr}_x(\sigma \xi), \qquad \xi\sim \mathcal N(0,\M^{-1}).
\end{equation}
For nonlinear retractions, this procedure induces a normalized pushforward Markov kernel rather than a flat Euclidean Gaussian density. 
Dropping the retraction Jacobian in local coordinates would necessitate a state-dependent reweighting term to ensure 
normalization. 
To bypass this explicit Jacobian computation, we avoid closed-form density evaluation in Experiments 1 and 2, relying solely on sampling and score matching. 
In the low-noise local-coordinate regime used here, the sampling-defined denoising target and the base-action loss are:
\begin{equation}
\label{eq:denoising_target}
t_\sigma(z,x) = \Pi_z\frac{x-z}{\sigma^2},
\end{equation}
\begin{equation}
\label{eq:base_action_loss}
\mathcal L_{S_0}(\theta) = \E_{x,\sigma,\xi} \left[ \frac{\sigma^2}{d} \left\|v_\theta(z,\sigma)-t_\sigma(z,x)\right\|_\M^2 \right],
\end{equation}
where \(\Pi_z\) is the tangent projection, \(d\) is the effective dimension, 
and \(\|v\|_\M^2=v^\top\M v\).

\subsection{Audit: Experiment 1}

In Experiment 1, we use a low-dimensional ALA2 ensemble to test whether our diffusion 
architecture can represent nontrivial base-action and operator structures.
Trajectories are obtained from a public dataset \cite{mcgibbon2014ala2}, with OpenMM providing the reference reduced-actions and forces \cite{eastman2017openmm}.

We first train a configuration-only, low-noise DSM model for \(S_0\). 
As detailed in Table~\ref{tab:s0_performance}, we evaluate the scaling performance across frame counts and pretraining settings, followed by a component-wise audit of the 256-frame checkpoint. These capacity-sensitive trends demonstrate successful (albeit uncalibrated) \(S_0\) recovery, even though the individual physical components are not equally well-recovered due to representation degeneracies. 
Having established the base-action audit, we next turn to ligand or condition
perturbations as operator fields.

\begin{table}[htbp]
\centering
\caption{Experiment 1A: Base action \(S_0\) recovery under scaling gates and component-head audit.}
\label{tab:s0_performance}
\small
\setlength{\tabcolsep}{4pt}
\begin{tabular*}{\linewidth}{@{\extracolsep{\fill}}lccc}
\toprule
\textbf{Config. / Component} &
\textbf{Score} &
\textbf{Action} &
\textbf{Action} \\
&
\textbf{NRMSE} &
\textbf{Pearson} &
\textbf{Aff. NRMSE} \\
\midrule
\multicolumn{4}{l}{\textit{\textbf{\(S_0\) scaling gate}}} \\
16 frames                         & 0.6392 & 0.3452 & 0.9359 \\
64 frames                         & 0.5814 & 0.6600 & 0.7474 \\
256 frames                        & 0.5802 & 0.7721 & 0.6354 \\
16 frames, 4096 pretrain          & 0.5532 & 0.4668 & 0.8833 \\
16 frames, larger model           & 0.6367 & 0.4570 & 0.8866 \\
\midrule
\multicolumn{4}{l}{\textit{\textbf{Component-head audit, 256-frame checkpoint}}} \\
Assembled total                   & -- & 0.780  & 0.621 \\
Bond (1--2)                       & -- & 0.759  & 0.629 \\
Angle (1--3)                      & -- & 0.920  & 0.392 \\
Torsion                           & -- & 0.612  & 0.790 \\
Direct                            & -- & $-0.175$ & 0.983 \\
Exception (1--4)                  & -- & 0.313  & 0.907 \\
Direct + exception                & -- & $-0.053$ & 0.998 \\
\bottomrule
\end{tabular*}
\end{table}

\section{Ligand perturbations as operator fields}

\subsection{Physical picture}

Attempting to learn the base action \(S_0(x)\) and the perturbation operator \(O(x)\) simultaneously via joint conditional score matching is ill-conditioned, as the unconstrained base action and perturbation operator readily absorb each other's representation errors. 
To resolve this degeneracy, we must declare or fix \(S_0(x)\) first. 
Once the base environment is fixed, a ligand perturbation can be represented as a systematic action shift along an alchemical path coordinate \(\kappa\):
\begin{equation}
\label{eq:perturbed_action}
S_\kappa(x)=S_0(x)+\kappa O(x).
\end{equation}

For noised variables, the unnormalized effective action \(S_{\kappa,\sigma}(z)\) satisfies:
\begin{equation}
\label{eq:noised_effective_action}
e^{-S_{\kappa,\sigma}(z)} = \int K_\sigma(z\mid x)e^{-S_0(x)-\kappa O(x)}\dd x.
\end{equation}
Differentiating Equation~\eqref{eq:noised_effective_action} with respect to \(\kappa\) under the integral yields the effective operator field:
\begin{equation}
\label{eq:effective_operator_field}
O_\sigma(\kappa,z) = \partial_\kappa S_{\kappa,\sigma}(z) = \E_{\kappa,\sigma}[O(x)\mid z].
\end{equation}
Assuming a \(\kappa\)-independent noising kernel, Equation~\eqref{eq:effective_operator_field} demonstrates that the effective operator field is diffusion-native, corresponding directly to the alchemical derivative of the noised effective action.

\subsection{Cross-entropy operator selection}

Condition labels provide a density-ratio route to scalar operator learning. 
For samples from a set of parameterized ensembles \(p_{\kappa_j}(x)\propto e^{-S_0(x)-\kappa_jO(x)}\), 
a Bayesian classifier yields logits of the form:
\begin{equation}
\label{eq:ce_logits}
\ell_j(x)=b_j-\kappa_j O_\theta(x),
\end{equation}
where \(b_j\) absorbs the class priors and free-energy constants. 
The cross-entropy loss is formulated as:
\begin{equation}
\label{eq:ce_loss}
\mathcal L_{\mathrm{CE}}(\theta,b) = -\E_{(x,y)} \log \frac{\exp[b_y-\kappa_yO_\theta(x)]}{\sum_j\exp[b_j-\kappa_jO_\theta(x)]}.
\end{equation}
For a binary endpoint pair (\(\kappa \in \{0, 1\}\)), this recovers a logistic density-ratio model:
\begin{equation}
\label{eq:binary_logistic_density}
\log \frac{p_1(x)}{p_0(x)} = -O(x)+\Delta F.
\end{equation}
Because cross-entropy is a classification objective providing only scalar-level density-ratio supervision, the resulting gradients of \(O_\theta\) do not reliably represent fine-grained microscopic forces.  
We therefore use cross-entropy primarily to locate a restricted operator basin, leaving the resolution of local forces to subsequent score matching.

\subsection{Clean score matching for operator fields}

With \(S_0\) fixed, the score of the perturbed action is:
\begin{equation}
\label{eq:perturbed_score_field}
v_\kappa(x) = -\M^{-1}\nabla S_\kappa(x) = v_0(x)-\kappa\M^{-1}\nabla O(x).
\end{equation}
Defining the operator score via automatic differentiation of the parameterized 
scalar operator, \(g_\theta(x)=-\M^{-1}\nabla O_\theta(x)\), we train \(O_\theta\) 
on clean samples from \(p_\kappa\). Dropping \(\theta\)-independent terms, the 
Hyvarinen score-matching objective for \(v_0+\kappa g_\theta\) is:
\begin{equation}
\label{eq:clean_score_matching_loss}
\mathcal L_{\mathrm{cleanSM}}(\theta) = \E_{p_\kappa} \left[ \kappa\langle g_\theta,v_0\rangle_\M +\frac{1}{2}\kappa^2\|g_\theta\|_\M^2 +\kappa\,\mathrm{div}_\M g_\theta \right],
\end{equation}
where the divergence is estimated via sliced finite differences in tangent directions. 
Crucially, fixing the base action \(S_0\) prevents mutual error absorption between 
\(S_0\) and \(O\), rendering the score-matching objective in Equation~\eqref{eq:clean_score_matching_loss} 
mathematically well-posed. To maintain global thermodynamic consistency, this local score-matching refinement is initialized or regularized within the restricted operator basin previously located by cross-entropy selection, preventing unconstrained global drift during force alignment.

\subsection{Audit: Experiment 2}

In Experiment 2, we evaluate whether the operator field can be recovered and audited once \(S_0\) is fixed. 
Experiment 2A uses a theoretical OpenMM \(S_0\) and applies cross-entropy to locate the scalar operator basin, followed by
clean on-manifold sliced score matching to refine the local operator score.
As shown in Table~\ref{tab:exp2_operator_recovery}, CE already recovers a strong held-out scalar operator signal, while CE+CSM preserves this scalar recovery and improves both operator-score and finite-difference gradient audits. 
Shuffled-\(\kappa\) CSM does not reproduce the force-level gain, and shuffled-CE collapses both scalar and local-gradient recovery. 
These results indicate that once \(S_0\) is anchored, operator-directed supervision can recover an auditable scalar \(O\) together with partial tangent-gradient information.

\begin{table}[htbp]
\centering
\caption{Experiment 2A: fixed-\(S_0\) operator recovery with cross-entropy (CE) and clean score matching (CSM). Values are mean \(\pm\) standard deviation over five seeds in the high-budget common-random audit.}
\label{tab:exp2_operator_recovery}
\small
\setlength{\tabcolsep}{3pt}
\begin{tabular*}{\linewidth}{@{\extracolsep{\fill}}lcccc}
\toprule
\textbf{Audit metric} &
\textbf{CE} &
\textbf{CE+CSM} &
\textbf{CE+shuf.-CSM} &
\textbf{shuf.-CE+CSM} \\
\midrule
Scalar \(O\) Pearson
& \(0.855\pm0.031\)
& \(0.830\pm0.045\)
& \(0.872\pm0.043\)
& \(0.219\pm0.118\) \\
Operator-score cosine
& \(0.360\pm0.079\)
& \(0.487\pm0.044\)
& \(0.358\pm0.082\)
& \(0.153\pm0.061\) \\
Operator-score NRMSE
& \(1.015\pm0.064\)
& \(0.905\pm0.025\)
& \(0.984\pm0.066\)
& \(1.051\pm0.053\) \\
Finite-difference Pearson
& \(0.405\pm0.094\)
& \(0.531\pm0.053\)
& \(0.409\pm0.098\)
& \(0.129\pm0.097\) \\
Finite-difference NRMSE
& \(0.909\pm0.042\)
& \(0.845\pm0.032\)
& \(0.906\pm0.048\)
& \(0.987\pm0.015\) \\
\bottomrule
\end{tabular*}
\end{table}

In this ALA2 audit, the operator \(O(x)\) is a smooth periodic function of selected backbone dihedral coordinates. 
The operator models are trained on biased ALA2 ensembles at
\(\kappa\in\{-2,-1,0,1,2\}\), using 320 configurations per training \(\kappa\). 
Evaluation is performed on held-out biased ensembles at
\(\kappa\in\{-1.5,-0.75,0.75,1.5\}\).

\section{A noisy reduced-potential bridge for free-energy readout}

\subsection{Why a noisy bridge can preserve free energy}

The preceding sections define the requisite semantics of \(S_0\) and \(O\). 
In this section, we demonstrate how to extract free energy from a diffusion model whether \(S\) is declared or learned.
For the protein--ligand endpoint studies below, we denote the two endpoint alchemical states by \(q=0\) and \(q=1\); this \(q\) is the binary endpoint specialization of the path coordinate \(\kappa\).
Free energy requires the partition function:
\begin{equation}
\label{eq:free_energy_partition}
F_\kappa=-\log Z_\kappa, \qquad Z_\kappa=\int e^{-S_\kappa(x)}\dd x.
\end{equation}
Because learned actions are identified only up to arbitrary additive constants, we must manually fix an action gauge to resolve the partition-function ambiguity. 
Even when the operator is known, endpoint coordinate ensembles often suffer from negligible phase-space overlap.
The key thermodynamic invariant of our formulation is that normalized forward noising preserves the partition function \(Z_\kappa\):
\begin{equation}
\label{eq:partition_invariance}
Z_{\kappa,\sigma} = \int e^{-S_{\kappa,\sigma}(z)}\dd \nu(z) = \iint K_\sigma(z\mid x)e^{-S_\kappa(x)}\dd x\,\dd\nu(z) = \int e^{-S_\kappa(x)}\dd x = Z_\kappa,
\end{equation}
which holds under the condition that the noising kernel is \(\kappa\)-independent and normalized over the noisy space, such that \(\int K_\sigma(z\mid x)\dd\nu(z)=1\) for every clean configuration \(x\) without state-dependent truncation or reweighting.

When the molecular coordinate space is reduced to a lower-dimensional feature space, we define the forward noising kernel \(K_\sigma\) and the reference measure directly on this reduced feature space. 
By using the empirical endpoint ensembles to define the pushforward measure, we completely bypass the need for a Cartesian-to-feature change-of-variables Jacobian. This feature-space formulation ensures that the partition function invariance \(Z_{\kappa,\sigma} = Z_\kappa\) holds exactly, provided a common normalized pushforward kernel is applied consistently to both endpoints. 
Consequently, we can write a density-ratio identity in noisy space that preserves the clean free energy:
\begin{equation}
\label{eq:noisy_density_ratio}
\log\frac{p_{1,\sigma}(z)}{p_{0,\sigma}(z)} = -\left[S_{1,\sigma}(z)-S_{0,\sigma}(z)\right]+\Delta F.
\end{equation}
Representing the action difference as a path integral of the effective operator:
\begin{equation}
\label{eq:quadrature_integral}
d^\sigma(z) = \int_0^1O_\sigma(\kappa,z)\dd\kappa = S_{1,\sigma}(z)-S_{0,\sigma}(z),
\end{equation}
we obtain the fundamental noisy density-ratio identity:
\begin{equation}
\label{eq:noisy_ratio_quadrature}
\log\frac{p_{1,\sigma}(z)}{p_{0,\sigma}(z)} = -d^\sigma(z)+\Delta F.
\end{equation}

\subsection{Shared-operator bridge objective}

To compute the quadrature in Equation~\eqref{eq:noisy_ratio_quadrature}, we train a sigma-conditioned operator network \(O_\theta(\kappa,z,\sigma) \approx O_\sigma(\kappa,z)\) alongside a shared scalar \(\Delta F_\theta\). 
The predicted action difference is computed via quadrature:
\begin{equation}
\label{eq:predicted_quadrature}
d^\sigma_\theta(z) = \int_0^1O_\theta(\kappa,z,\sigma)\dd\kappa.
\end{equation}
The unified training objective minimizes four terms:
\begin{equation}
\label{eq:total_bridge_loss}
\mathcal L_{\mathrm{bridge}} = w_O\mathcal L_O + w_A\mathcal L_A + w_R\mathcal L_R + w_S\mathcal L_S.
\end{equation}
The first term, \(\mathcal L_O\), regresses the endpoint operator evaluations when physical targets \(\widehat O_\sigma\) are available (as in supervised settings):
\begin{equation}
\label{eq:loss_o}
\mathcal L_O = \E\left[ \left(O_\theta(\kappa,z,\sigma)-\widehat O_\sigma\right)^2 \right].
\end{equation}
To fix the additive gauge of \(d^\sigma_\theta\) at the zero-noise limit (\(\sigma \to 0^+\)), we introduce the clean action-integral anchor \(\mathcal L_A\):
\begin{equation}
\label{eq:loss_a}
\mathcal L_A = \E_x \left[ \left( \int_0^1O_\theta(\kappa,x,0^+)\dd\kappa -\Delta S(x) \right)^2 \right], \qquad \Delta S(x)=S_1(x)-S_0(x),
\end{equation}
which evaluates the direct-path action difference \(\Delta S(x)\) on clean endpoint features with a near-zero sigma label for numerical stability. 

In unsupervised coordinate-only learning where analytical potentials are absent, this shift degeneracy is alternatively resolved by fixing an explicit action convention (such as centering \(O_\theta\) on the source training pool). 
Without such an anchor or centering convention, \(d^\sigma_\theta\) and \(\Delta F_\theta\) retain a shift degeneracy even if the endpoint operator and density-ratio consistency are otherwise accurate.

The third term, \(\mathcal L_R\), enforces density-ratio consistency using the noisy endpoint identity:
\begin{equation}
\label{eq:loss_r}
\mathcal L_R = \E_{z,y} \left[ \mathrm{BCE} \left( \Delta F_\theta-d^\sigma_\theta(z),y \right) \right],
\end{equation}
where \(y=0\) for noisy source endpoint samples, \(y=1\) for noisy target endpoint samples, and training uses balanced batches to prevent the logistic intercept from absorbing unbalanced class priors. Finally, \(\mathcal L_S\) is a mild smoothness regularizer in \(\kappa\). The scalar \(\Delta F_\theta\) is learned directly by the bridge without relying on intermediate-\(\kappa\) ensembles, nonequilibrium trajectories, or accept/reject steps.

\subsection{Controlled study: Experiment 3}

Experiment 3 transfers our action-operator semantics to a controlled 185L/IND protein--ligand bound-state system based on PDB entry 185L, utilizing OpenMM and OpenFF-family small-molecule parameterizations \cite{pdb185l,morton1995,eastman2017openmm,mobley2018smirnoff,openff2024sage}.

We first introduce a physical C6-H to C6-F ligand-atom mutation. Importantly, this operator is not a full force-field replacement or a complete ligand mutation free energy. It is a restricted single-topology ligand-pocket nonbonded perturbation: \(O(x)\) is defined as the change in nonbonded interaction energy between the IND ligand atoms and a fixed set of pocket heavy atoms. 
Due to negligible phase-space overlap between the endpoints, direct endpoint FEP fails, yielding a bidirectional forward-reverse discrepancy (hysteresis) of 28 to 33 \(\kBT\).
An independent 19-state split-path MBAR calculation provides a stable, converged validation reference of \(-25.8219\ \kBT\) for subsequent comparison. 
We then implement the main noisy operator bridge, utilizing only the endpoint states (\(q=0, 1\)) along with per-frame evaluations of \(S_0(x)\) and the direct-path action difference \(\Delta S(x)\) to supervise the clean action-integral anchor \(\mathcal{L}_A\). 
In this setup, \(\Delta S\) is withheld from the input features and acts solely as the zero-noise target to fix the additive gauge.

\begin{table}[htbp]
\centering
\caption{Experiment 3: Free-energy estimates and high-noise diagnostics for the 185L/IND C6-H to C6-F mutation. All values are in units of \(k_{\mathrm B}T\). Errors are absolute errors relative to the 19-state split-path MBAR reference.}
\label{tab:exp3_results}
\small
\setlength{\tabcolsep}{5pt}
\begin{tabular*}{\linewidth}{@{\extracolsep{\fill}}lcc}
\toprule
\textbf{Method / Configuration} &
\textbf{Estimate} &
\textbf{Abs. Error} \\
\midrule
19-state split-path MBAR reference
& \(-25.8219\)
& -- \\
\midrule
\textbf{Noisy operator bridge, 6 seeds}
& \(\mathbf{-24.7405\pm0.0952}\)
& \(\mathbf{1.0814\pm0.0952}\) \\
\quad high-noise BAR layer, \(\sigma=0.03\)
& \(-24.7041\pm0.1685\)
& \(1.1178\pm0.1685\) \\
\quad high-noise BAR layer, \(\sigma=0.04\)
& \(-24.6877\pm0.1856\)
& \(1.1341\pm0.1856\) \\
\bottomrule
\end{tabular*}
\end{table}

As summarized in Table~\ref{tab:exp3_results}, the main bridge yields a shared free-energy estimate \(\Delta F_\theta = -24.7405 \pm 0.0952\ \kBT\), matching the stable 19-state reference within \(1.0814 \pm 0.0952\ \kBT\).

\subsection{Endpoint-only gauge audit: Experiment 4}

Experiment 4 addresses a more difficult question: can we identify free energies from endpoint coordinates and labels alone? 
To address this, we apply our endpoint-only audit to the 185L/IND protein--ligand bound state, where the base action \(S_0\) corresponds to the unmutated (\(q=0\)) pocket environment, and the operator \(O\) represents either the C6-H3 to F or the IND C4-H7 to F ligand-atom mutation.
The protocol first trains and freezes a \(q=0\) base representation \(S_{0,\theta}\) via feature-space denoising score matching. 
It then learns a scalar operator \(O_\theta\) purely from balanced \(q=0/1\) endpoint ensembles.

Because the learned operator has an arbitrary additive gauge, we fix its constant offset by centering the predictions on the \(q=0\) training pool:
\begin{equation}
\label{eq:centering_gauge}
O_\theta(x) = \tilde{O}_\theta(x) - \E_{x \in \mathrm{train}, q=0} [\tilde{O}_\theta(x)].
\end{equation}
The corresponding audit target is thus the \(q=0\)-centered physical scale, verifying whether coordinate distributions alone can recover the nonconstant operator shape.

We evaluate this unsupervised audit on both the original C6-H3 to F mutation and a second single-topology mutation (IND C4-H7 to C4-F7) to rule out single-site bias. 
As summarized in Table~\ref{tab:exp4a_results}, the main model successfully recovers the \(q=0\)-centered operator scale for both sites, achieving endpoint-pool audit errors of \(0.8149 \pm 0.3496\ \kBT\) for C6 and \(0.7324 \pm 0.0747\ \kBT\) for C4. 
In contrast, all control variants fail: shuffling labels collapses the BAR estimates to near-zero, while removing the \(q=0\) centering or the endpoint thermodynamic consistency constraints severely inflates the audit errors.

To establish a rigorous thermodynamic baseline, we generate an independent 19-state split-path MBAR reference for the C4 mutation, yielding a converged absolute free energy of \(-19.0432\ \kBT\).
Translated to the \(q=0\)-centered gauge, this corresponds to a reference value of \(-3.5935 \pm 0.1651\ \kBT\). Against this independent split-MBAR-centered reference, our coordinate-only estimator predicts the scale with an error of \(1.3676 \pm 0.0747\ \kBT\).

\begin{table}[htbp]
\centering
\caption{Experiment 4: endpoint-only gauge audit for C6 and C4 mutations. Values are in units of \(k_{\mathrm B}T\) and reported as mean \(\pm\) standard deviation over five seeds. Errors are evaluated against the endpoint-pool \(q=0\)-centered audit reference.}
\label{tab:exp4a_results}
\small
\setlength{\tabcolsep}{3pt}
\begin{tabular*}{\linewidth}{@{\extracolsep{\fill}}lcccc}
\toprule
\textbf{Method / Control}
& \multicolumn{2}{c}{\textbf{C6-H3 \(\to\) F}}
& \multicolumn{2}{c}{\textbf{C4-H7 \(\to\) F}} \\
\cmidrule(lr){2-3} \cmidrule(lr){4-5}
& \textbf{BAR} & \textbf{Err.}
& \textbf{BAR} & \textbf{Err.} \\
\midrule
\textbf{CE + TC}
& \(\mathbf{-2.2419\pm0.1125}\)
& \(\mathbf{0.8149\pm0.3496}\)
& \(\mathbf{-2.2259\pm0.1513}\)
& \(\mathbf{0.7324\pm0.0747}\) \\
Shuffled labels
& \(-0.0002\pm0.0013\)
& \(3.0565\pm0.4416\)
& \(0.0011\pm0.0044\)
& \(1.4946\pm0.1652\) \\
No \(q=0\) centering
& \(-0.1182\pm0.5476\)
& \(2.9386\pm0.3890\)
& \(0.0620\pm0.4313\)
& \(1.5555\pm0.3228\) \\
No endpoint TC
& \(-8.1991\pm3.4096\)
& \(5.1423\pm3.1922\)
& \(-5.8005\pm2.4880\)
& \(4.3070\pm2.3313\) \\
\bottomrule
\end{tabular*}
\end{table}

These unsupervised results strengthen the semantic narrative of this work. 
They confirm that endpoint coordinates can support a useful gauge-fixed free-energy readout when thermodynamic consistency is enforced.

\section{Discussion and conclusion}

This work establishes a physically rigorous action-operator semantics for molecular diffusion models, enabling the direct readout and audit of internalized thermodynamic quantities such as base actions (\(S_0\)), perturbation operators (\(O\)), and free energy differences (\(\Delta F\)). 
This framework successfully extracts absolute free energy differences in a supervised protein--ligand mutation (Experiment 3) and uncovers relative free energy scales purely from endpoint coordinates and labels in an unsupervised audit (Experiment 4).

While this work provides a foundational language to interpret and audit diffusion models as thermodynamic estimators, significant challenges remain for real-world pharmaceutical applications. 
First, fully robust free energy calculation in drug discovery often requires evaluating \textbf{relative free energy differences (\(\Delta\Delta F\))} between multiple ligands. 
This necessitates solving the more complex problem of a universally fixed thermodynamic gauge across diverse chemical transformations, moving beyond our current system-specific or endpoint-centered conventions. 
Second, while our controlled studies confirm the identifiability of \(S_0\), \(O\), and \(F\) within a specific framework, it remains crucial to validate whether these physically interpretable quantities can be accurately and robustly extracted in \textbf{truly large-scale generative models} trained on vast and diverse biomolecular datasets.

The ultimate goal is to evolve current black-box diffusion models into auditable thermodynamic machines capable of predicting drug-like properties. 
Achieving this requires scaling the joint recovery of \(S_0\), condition-dependent \(O\), and \(O_\sigma(\kappa,z)\) across numerous protein--ligand environments, guided by the robust semantic framework laid out in this work. 
This represents a rigorous path toward transforming generative molecular models into powerful, interpretable tools for chemical and biological discovery.

\section*{Acknowledgments}
This work was supported by Research Grants Council of Hong Kong (CRF C7015-24G, GRF 17311322
and CRF C7012-21GF) and National Natural Science Foundation of China (Grant
No. 12222416).

\bibliographystyle{naturemag}
\bibliography{ref}

\end{document}